\documentstyle[12pt]{article}

\newcommand{\be}{\begin{equation}}
\newcommand{\ee}{\end{equation}}

\begin{document}

\begin{center}

{\Large{\bf Coherent Optics and Localized Light} \\ [5mm]

V.I. Yukalov} \\ [3mm]

{\it Centre for Interdisciplinary Studies in Chemical Physics \\
University of Western Ontario, London, Ontario N6A 3K7, Canada \\
and \\
Bogolubov Laboratory of Theoretical Physics \\
Joint Institute for Nuclear Research, Dubna 141980, Russia}

\end{center}

\vspace{3cm}

\begin{abstract}

Interplay between the effects of coherent radiation and localization of
light is analysed. A system of two--level atoms is placed in a medium
interacting with electromagnetic field. The matter--light interaction can
result in the appearance of a band gap in the spectrum of polariton states. 
If an atom with a resonance frequency inside the gap is incorporated into 
such a medium, the atomic spontaneous emission is suppressed, which is
termed the localization of light. However, a system of resonance atoms
inside the gap can radiate due to their coherent interactions. The
peculiarity of the coherent radiation by a system of atoms, under the
localization of light for a single atom, is studied.

\end{abstract}

\newpage

\section{Localization of Light}

In some cases a medium may possess an electromagnetic band gap where a
severe depressions of the photon density of states occurs. One such case
is rather well known -- This is the appearance of the polariton band gap
due to the interaction of light with a dense medium [1-3]. Another
possibility has been recently found when a spectral photon gap develops
because of periodicity of dielectric structures [4,5]. The latter are
called photonic band--gap materials. In such dielectric superlattices,
strong localization of photons happens [4-6]. The same effect of light
localization arises if a resonance atom, with a frequency inside the
polariton band gap, is doped into a dispersive medium [7-9]. If the atomic
resonance frequency lies near the gap, a polariton--atom bound state
appears with an eigenfrequency lying within the gap. The appearance of
this bound state results in a significant suppression of spontaneous
emission, that is, in localization of light. This means that an atom, in
the stationary state, has a finite probability to be in the excited state,
provided that it was excited at the initial moment of time.

To explain in simple parlance what does mean the localization of light,
let us consider a two--level atom whose population difference is described
by an operator $\sigma^z(t)$. It is convenient to introduce the excitation
function
$$ \eta(t)\equiv\frac{1}{2}\left ( 1 +s(t)\right ) , \qquad
s(t)\equiv\langle\sigma^z(t)\rangle , $$
in which the angle brackets $\langle\ldots\rangle$ imply a quatum--mechanical
averaging. The atom is excited when the population difference $s=1$, i.e.,
the excitation function $\eta=1$. When the atom is not excited, then
$s=-1$ and $\eta=0$. There exists a finite probability to find the atom
excited, in the stationary state, if and only if
$$ \lim_{t\rightarrow\infty}\eta(t)\neq 0 , \qquad
\lim_{t\rightarrow\infty}s(t) =\zeta\neq -1 . $$
This is exactly what one means under the {\it localization of light}. The
reason for this localization is the suppression of spontaneous emission,
if the atom was initially excited. If the suppression is absolute, then,
starting with $s(0)=1$, one comes to $\zeta=1$, that is, the atom does not
become deexcited, keeping an absorbed photon forever. Vice versa, if an
atom, with a frequency inside the polariton band gap, was not excited at
the initial time $t=0$, that is, $s(0)=-1$, then it should remain not
excited in the stationary state, as $t\rightarrow\infty$, since there are
no photons in the gap, which could excite the atom. The corresponding
dynamics of the population difference can be described by the equation
$$ 
\frac{ds}{dt} =-\gamma_1( s-\zeta ) , \qquad \zeta\approx s(0) , 
$$
whose evident solution $s(t)=\zeta\approx s(0)$ demonstrates the
suppression of spontaneous emission.

The situation becomes more complicated when a collection of identical
impurity atoms, with a transition frequency in the polariton band gap, is
incorporated into the medium. If the spacing between the admixture atoms
is much smaller than the transition wavelength, then the electromagnetic
coupling of atoms leads to the formation of a photonic impurity band
within the polarization band gap [10]. If the density of the admixture is
high, the polariton gap can be destroyed at all. Electromagnetic field can
propagate in the impurity band formed by collective interactions of atoms,
and coherent radiation becomes possible. If the spacing between the
admixture atoms would be much larger than the transition wavelength, then
the propagation band could not be formed and the atomic radiation would be
prohibited. In this way, only coherent interactions can overcome the
suppression of emission caused by the localization of light.

The time evolution of spontaneous emission near the edge of a photonic
band gap has been considered [11,12] for a simple concentrated Dicke
model, where the radiation wavelength is assumed to be much larger than
not only the interatomic spacing but the whole system. This model,
evidently, is equivalent to a single--atom model with atomic variables
factored by the number of atoms.

The aim of the present communication is to suggest a more realistic,
though yet solvable, approach to describing coherent emission of admixture
atoms placed in a medium with localizing light. Clearly, a more realistic
approach is, at the same time, more and even much more complicated.
Therefore, it would be impossible in the frame of this communication to
expound it in whole. The main attention here is paid to the the formulation 
of the problem, with a brief survey of physical picture, and some new
results are announced.

\section{Formulation of Problem}

Consider a system of $N$ resonance two--level radiators enumerated by the
index $i=1,2,\ldots,N$. These can be atoms, molecules, nuclei, or quantum
dots. For short, let us call them atoms. Their Hamiltonian is
\be
\hat H_a =\frac{1}{2}\sum_{i=1}^N \omega_0 (1 +\sigma_i^z ) ,
\ee
where $\omega_0$ is a transition frequency $(\hbar\equiv 1)$ and
$\sigma_i^z$ is a population difference operator. The electromagnetic
field Hamiltonian has the general form
\be
\hat H_f = \frac{1}{8\pi} \int \left [ \stackrel{\rightarrow}{E}^2
(\stackrel{\rightarrow}{r}) + \stackrel{\rightarrow}{H}^2
(\stackrel{\rightarrow}{r}) \right ] d\stackrel{\rightarrow}{r} ,
\ee
with the electric field $\stackrel{\rightarrow}{E}$ and magnetic field
$\stackrel{\rightarrow}{H}\equiv\stackrel{\rightarrow}{\nabla}\times
\stackrel{\rightarrow}{A}$, where $\stackrel{\rightarrow}{A}$ is the
vector--potential satisfying the Coulomb gauge condition 
$\stackrel{\rightarrow}{\nabla}\cdot\stackrel{\rightarrow}{A}=0$. The
interaction between the atoms and field, in the dipole approximation, is
given by
\be
\hat H_{af} = -\frac{1}{c}\sum_{i=1}^N\stackrel{\rightarrow}{J}_a
(\stackrel{\rightarrow}{r}_i)\stackrel{\rightarrow}{A}
(\stackrel{\rightarrow}{r}_i) ,
\ee
with the transition current
\be
\stackrel{\rightarrow}{J}_a(\stackrel{\rightarrow}{r}_i) = 
i\omega_0\left ( \sigma_i^+\stackrel{\rightarrow}{d}^* -
\sigma_i^-\stackrel{\rightarrow}{d} \right ) ,
\ee
in which $\sigma_i^\pm$ is a raising or lowering operator, respectively;
and $\stackrel{\rightarrow}{d}$, a transition dipole. As usual [13], the
relativistic term $\stackrel{\rightarrow}{A}^2/c^2$ is neglected. The
system of radiators defined by Eqs.(1) to (4) forms the basis for the
standard consideration of collective processes in the emission of light
[14,15].

The case we are considering here is aggravated by the fact that the
resonance atoms are not in empty space but are inserted as admixtures into
a medium. The latter can be modeled in different ways, with the main
requirement that a band gap should appear resulting in the localization of
light for a single atom. Photonic band--gap materials can be described by
periodic superstructures of scatterers [4,5]. A frequency gap for
propagating electromagnetic modes exists also in many natural dielectrics
and semiconductors [1,2]. For instance, the polariton effect is well
developed in such semiconductors as $CuCl,\; CuBr,\; CdS,\; CdSe,\;
ZnSe,\; GaAs,\; GaSb,\; InAs,\; AlAs,\; SiC$, and in some semiconductor
microstructures including quantum dots, wells, and wires [16]. A frequency
gap for light propagation inside dense media is known to appear when the
latter contain excitations being in resonance with the frequency of light
[1,2]. For example, the matter could be presented as an ensemble of
two--level radiators. Then, these radiators together with the admixture
atoms would form a kind of a resonance two--component system [17-19]. The
polariton gap in dispersive dense media arises because of the interaction
of light with some gapful elementary excitations, like excitons or optical
phonons [1,2]. Hence, a medium can be modeled by an ensemble of oscillators, 
possessing an optical branch, which represent optical--type collective
excitations of the medium. A Hamiltonian of such collective excitations
has, in general, the form
\be
\hat H_m =\sum_{j=1}^{N'}\frac{\stackrel{\rightarrow}{p}_j^2}{2m}
+\frac{1}{2}\sum_{ij}^{N'}\sum_{\alpha\beta}^3 D_{ij}^{\alpha\beta}
u_i^\alpha u_j^\beta ,
\ee
where $N'$ is the number of lattice sites; $\stackrel{\rightarrow}{p}_i$
and $\stackrel{\rightarrow}{u}_i$ are momentum and displacement operators,
respectively; $D_{ij}^{\alpha\beta}$ is a dynamical matrix. The interaction
of the matter excitations with electromagnetic field is described by the
term
\be
\hat H_{mf} =-\frac{1}{c}\sum_{j=1}^{N'}\stackrel{\rightarrow}{J}_m
(\stackrel{\rightarrow}{r}_j)\stackrel{\rightarrow}{A}
(\stackrel{\rightarrow}{r}_j) ,
\ee
in which
\be
\stackrel{\rightarrow}{J}_m(\stackrel{\rightarrow}{r}_j) =
\frac{e}{m}\stackrel{\rightarrow}{p}_j 
\ee
is a local current at the point $\stackrel{\rightarrow}{r}_j$ of the medium.

In this way, the total Hamiltonian of the considered system is
\be
\hat H = \hat H_a + \hat H_f + \hat H_{af} + \hat H_m +\hat H_{mf} ,
\ee
consisting of the atom Hamiltonian (1), field term (2), atom--field
interaction (3), medium Hamiltonian (5), and of the medium--field interaction
(6).

The stationary states of the system with Hamiltonian (8) can be studied as
follows. One expands the vector potential
\be
\stackrel{\rightarrow}{A} =
\sum_{k\nu}\left (\frac{2\pi c}{kV}\right )^{1/2} \left (
a_{k\nu}\stackrel{\rightarrow}{e}_{k\nu}e^{i\stackrel{\rightarrow}{k}
\stackrel{\rightarrow}{r}} + a^\dagger_{k\nu}
\stackrel{\rightarrow}{e}_{k\nu}^*e^{-i\stackrel{\rightarrow}{k}
\stackrel{\rightarrow}{r}}\right )
\ee
in plane waves, with $\stackrel{\rightarrow}{e}_{k\nu}$ being a polarization 
vector; $k\equiv|\stackrel{\rightarrow}{k}|;\; \stackrel{\rightarrow}{k}$,
a wave vector; $\nu=1,2$; $V$, volume; and $a_{k\nu}$ being a photon
operator indexed by the wave vector $\stackrel{\rightarrow}{k}$ and the
polarization--branch number $\nu$. The displacement and momentum operators 
of the medium are also expanded in plane waves:
$$ \stackrel{\rightarrow}{u}_j =
\sum_{ks}\left ( 2mN'\omega_{ks}\right )^{-1/2}\left ( b_{ks} +
b_{-ks}^\dagger\right )\stackrel{\rightarrow}{e}_{ks}
e^{i\stackrel{\rightarrow}{k}\stackrel{\rightarrow}{r}_j} , $$
\be
\stackrel{\rightarrow}{p}_j = - i
\sum_{ks}\left (\frac{m\omega_{ks}}{2N'}\right )^{1/2}\left ( b_{ks} -
b_{-ks}^\dagger\right )\stackrel{\rightarrow}{e}_{ks}
e^{i\stackrel{\rightarrow}{k} \stackrel{\rightarrow}{r}_j} ,
\ee
where $\stackrel{\rightarrow}{e}_{ks}$ is a corresponding polarization
vector; $s=1,2,3$; and $\omega_{ks}$ is the spectrum of collective
excitations defined by the eigenvalue problem
\be
\frac{1}{mN'}\sum_{ij}^{N'}\sum_{\beta=1}^3 D_{ij}^{\alpha\beta}
e^{i\stackrel{\rightarrow}{k}\stackrel{\rightarrow}{r}_{ij}}e^\beta_{ks}=
\omega_{ks}^2e^\alpha_{ks} ,
\ee
with $\stackrel{\rightarrow}{r}_{ij}=\stackrel{\rightarrow}{r}_i -
\stackrel{\rightarrow}{r}_j$. The frequency and polarization vectors are
assumed to be even functions of the wave vector, $\omega_{ks}=\omega_{-ks}$,
$\stackrel{\rightarrow}{e}_{ks}=\stackrel{\rightarrow}{e}_{-ks}$. The
destruction, $b_{ks}$, and creation, $b_{ks}^\dagger$, operators of 
collective oscillations satisfy the Bose commutation relations. With
expansion (9), the field Hamiltonian (2) takes the known simple form
\be
\hat H_f =\sum_{k\nu} ck\left ( a_{k\nu}^\dagger a_{k\nu} +
\frac{1}{2}\right ) .
\ee
And the matter Hamiltonian (5), by means of (10), becomes
\be
\hat H_m =\sum_{ks}\omega_{ks}\left ( b_{ks}^\dagger b_{ks} +
\frac{1}{2}\right ) .
\ee
Recall that an optical--type spectrum $\omega_{ks}$ is to be assumed in
(13). Introducing polariton operators that are linear combinations of the
photon operators $a_{k\nu}$ and of the boson operators $b_{k\nu}$, one
can, in some cases [1,2], diagonalize the sum of the Hamiltonians 
$\hat H_f + \hat H_m +\hat H_{mf}$, obtaining a diagonal polariton
Hamiltonian. For example, a detailed description of this procedure of 
diagonalization can be found in Ref.[20], where a uniform and isotropic
model is considered and several simplifications, in line with the
Heitler--London and resonance approximation, are involved. In these
approximations, one neglects the counter--rotating terms and two--boson
transitions, whose influence, similarly to two--atom transitions [21], can
become important only far from the resonance.

When there is only one admixture atom in the medium, that is $N=1$, then
the stationary states of Hamiltonian (8) can be found [9,20] resorting to
the uniform, isotropic, and resonance approximations. If the atomic
resonance frequency lies near the gap, there appears a polariton--atom
bound state with an eigenfrequency lying within the gap. This means that
light is localized at the atom. As a result of the appearance of this
localized bound state, a significant suppression of spontaneous emission
occurs [9]. The behavior of many--polariton states is a little more
diverse. Those states containing an even number of polaritons correspond
to solitons that can propagate within the gap, while the states with an
odd number of polaritons represent solitons that are pinned to the atom
forming a many--polariton bound state. The latter state, similarly to the
single--polariton bound state, also depicts the localization of light [20].

When two identical two--level atoms are placed in a frequency dispersive
medium whose polariton spectrum has a gap, then the polariton--atom bound
state lying within the polaritonic gap splits into a doublet due to an
effective atom--atom interaction [22]. The spontaneous emission can exist
only if the resonance frequency of these two atoms lies in the polariton
continuous spectrum. And if the resonance frequency lies within the gap, then 
two discrete modes represent a doublet of bound polariton--atom states, for 
which spontaneous emission is practically completely suppressed [22,23].

A qualitatively different situation develops when many resonance admixture 
atoms are placed in the medium. Stationary states for this case have been
studied for a one--dimensional atomic chain incorporated in a uniform and
isotropic system [22,24]. The nearest--neighbor approximation has been
used. In the case of spatially correlated atoms, with a frequency inside
the gap, a polariton--impurity band is formed within the polaritonic gap. 
Then polaritons can propagate in this impurity band and the atomic chain
provides a waveguide for the radiation field. Even in the nearest--neighbor 
approximation, when, for a chain, the interaction of only three atoms is
effectively taken into account, the width of the impurity band, normalized
with respect to the polariton gap, is $0.14$ [24]. Since this width is,
roughly speaking, proportional to $N-1$, the collective interaction of
about $10$ atoms should fill the whole polariton gap.

The energy--momentum representation employed for studying stationary states 
is not convenient for considering space--time dynamics of collective 
processes. For the latter purpose, to our mind, it is more appropriate to
remain in real space and time. One may write the Heisenberg evolution
equations for the operators of the problem. Then, formally solving the
Maxwell equations, one may exclude the field variables (see details in
[25]). After that, one comes to the equations for the atomic variables,
$$ 
\frac{d\sigma_i^-}{dt} = - ( i\omega_0 +\gamma_2 )\sigma_i^- +
\sigma_i^z \stackrel{\rightarrow}{d}^*\cdot \stackrel{\rightarrow}{D}_i +
$$
\be
+
ik_0^2\sigma_i^z\stackrel{\rightarrow}{d}\cdot \sum_{j(\neq i)}^N
\frac{1}{r_{ij}}\left [ \sigma_j^+ \left ( t -\frac{r_{ij}}{c}\right )
\stackrel{\rightarrow}{d}^* -\sigma_j^- \left ( t-\frac{r_{ij}}{c}
\right )\stackrel{\rightarrow}{d}\right ] ,
\ee
$$
\frac{d\sigma_i^z}{dt} =-\gamma_1\left (\sigma_i^z -\zeta\right ) - 2
\left (\sigma_i^+\stackrel{\rightarrow}{d}^* + \sigma_i^-
\stackrel{\rightarrow}{d}\right )\cdot\stackrel{\rightarrow}{D}_i -
$$
\be
-2ik_0^2\left ( \sigma_i^+\stackrel{\rightarrow}{d}^* + \sigma_i^-
\stackrel{\rightarrow}{d}\right )\sum_{j(\neq i)}^N\frac{1}{r_{ij}}\left [
\sigma_j^+\left ( t -\frac{r_{ij}}{c}\right )\stackrel{\rightarrow}{d}^*
- \sigma_j^-\left ( t -\frac{r_{ij}}{c}\right )
\stackrel{\rightarrow}{d}\right ] ,
\ee
where $\gamma_1$ and $\gamma_2$ are the longitudinal and transverse
relaxation parameters, respectively; $k_0\equiv \omega_0/c$; 
$r_{ij}\equiv|\stackrel{\rightarrow}{r}_{ij}|$; and
$$ \stackrel{\rightarrow}{D}_i(t) =\frac{k_0}{c}
\sum_{j(\neq i)}^{N'}\frac{1}{r_{ij}}\stackrel{\rightarrow}{J}_m
\left (\stackrel{\rightarrow}{r}_j, t-\frac{r_{ij}}{c}\right ) $$
is an electric field produced by the medium. Assuming that the single--atom 
stationary state corresponds to localized light, we put $\zeta=\langle
\sigma_i^z(0)\rangle$, where $\langle\ldots\rangle$ means the statistical
averaging over an initial state. Equations (14) and (15) are complemented
by initial conditions
$$ u_0=\langle\sigma_i^-(0)\rangle , \qquad
s_0 = \langle\sigma_i^z(0)\rangle . $$

\section{New Results}

The notion of light localization is relatively recent. This is why the main
part of this communication has been devoted to the description of the 
related physical picture and to the formulation of the basic equations 
that would permit one to consider the time development of collective
phenomena for a system of admixture atoms in a medium with localized light.
The limited frames of this communication do not allow to expound in detail
the way of solving the basic equations (14) and (15) and the analysis of
the corresponding solutions. This will be done in a separate publication.
Here, we only can briefly delineate the scheme of solving Eqs.(14) and (15)
and present some fresh results.

From the operator equations (14) and (15), we pass to the equations for
the related statistical averages. The pair correlation functions are
decoupled in the semiclassical approximation. The retardation is treated
in the quasirelativistic approximation as in Refs. [26-28]. The system of
nonlinear differential equations is solved by means of the scale separation 
approach [29,30]. The resulting physical picture depends on the values of
the following parameters: initial conditions $u_0$ and $s_0$; the coupling
parameter of coherent atomic interactions,
\be
g =\frac{k_0^3d_0^2}{\gamma_2}\sum_{j(\neq i)}^N 
\frac{\sin(k_0r_{ij})}{k_0r_{ij}}\; ;
\ee
the effective parameter of coupling between the admixture atoms and
matter,
\be
\alpha = \langle\langle\left | e^{-\Gamma t}
\int_0^t e^{(i\Omega +\Gamma)\tau}\stackrel{\rightarrow}{d}^*\cdot
\stackrel{\rightarrow}{D}(\tau)d\tau\right |^2\rangle\rangle ,
\ee
where the double brackets $\langle\langle\ldots\rangle\rangle$ imply the
statistical and time averaging, and
\be
\Omega = \omega_0 +\Delta_L s , \qquad \Gamma =\gamma_2 ( 1 - gs)
\ee
are an effective frequency and attenuation, with
$$ \Delta_L = k_0^3d_0^2\sum_{j(\neq i)}^N
\frac{\cos(k_0r_{ij})}{k_0r_{ij}} ,
\qquad s = \langle \sigma_i^z\rangle ; $$
and the critical atom--matter coupling parameter
\be
\alpha_c =\frac{(1-gs_0)^2+4g^2|u_0|^2}{4g^2s_0^2}\; .
\ee

The overall physical picture for the case $g\gg 1$ and $\alpha\ll\alpha_c$
is as follows. If the admixture atoms at the initial time $t=0$ are excited, 
then after a delay time $t_0\ll T_1$, defined by the values of the above
parameters, a coherent burst occurs. Then, a series of coherent bursts
follows, separated from each other by the periods of practically no
radiations. This series of bursts lasts for the time of several $T_1$.
Finally, dynamics tends to a stationary state with the excitation function
\be
\lim_{t\rightarrow\infty}\eta(t) = \frac{1}{2}\left ( 1 +\frac{1}{g}
\right ) .
\ee
Remembering the definition in Section 1, we see that the limit (20)
exhibits a partial localization of light.

\vspace{5mm}

{\bf Acknowledgements}

\vspace{2mm}

I am grateful for discussions to M.R. Singh and W. Lau. A Senior Fellowship 
from the University of Western Ontario, Canada, is appreciated.

\end{document}